\documentclass[a4paper, 12pt]{article}
\usepackage{amsfonts}
\usepackage{amsmath}
\usepackage{harvard}

\begin{document}

\title{Rayleigh waves in symmetry planes of crystals: 
explicit secular equations \\and some explicit  wave speeds.}

\author{Michel Destrade}
\date{2003}
\maketitle

\bigskip

%
%
%
\begin{abstract}

Rayleigh waves are considered for crystals possessing 
at least one plane of symmetry. 
The secular equation is established explicitly 
for surface waves propagating in any direction of the plane of 
symmetry, using two different methods.
This equation is a quartic for the squared wave speed in general,
and a biquadratic for certain directions in certain crystals, where it 
may itself be solved explicitly.
Examples of such materials and directions are found in the case of
monoclinic crystals with the plane of symmetry at $x_3=0$.
The cases of orthorhombic materials and of incompressible materials
are also treated.

\end{abstract}

\newpage


\section{Introduction}

The simplest physical setting involving a boundary value
problem for an elastic material is that of a semi-infinite body with a
plane boundary left free of tractions.
However, the consideration of small amplitude deformations and motions
of such a half-space leads usually to considerable mathematical
difficulties, especially when the material is anisotropic.
Indeed in the case of general (triclinic) anisotropy, the equations
of motion (or of equilibrium) lead to the resolution of a sextic
for the partial inhomogeneous plane waves (or deformations),
whose roots cannot be obtained explicitly.
Consequently, closed-form solutions have been sought for materials
with at least orthorhombic symmetry, because then the equations of
motion lead to a biquadratic, and because this case covers 16
different types of common symmetry classes such as tetragonal,
hexagonal, or cubic \cite{RoDi84}.

In between the classes of triclinic  crystals (no plane of symmetry)
and of orthorhombic crystals (three orthogonal planes of symmetry),
is the class of monoclinic crystals, with only one plane of symmetry.
Among the three possibilities for the orientation of the symmetry
plane, the configuration of a half-space $x_2 \ge 0$ made of
monoclinic material with the plane of symmetry at $x_3 = 0$
is particularly important for two-dimensional deformations,
(a) because in-plane stress and in-plane strain decouple from 
anti-plane stress and anti-plane strain, respectively, so that the 
equations of motion yield a quartic; and
(b) because these materials are structurally invariant \cite{Ting00}
that is, the stress-strain relationships retain their form with respect
to rotations in the ($x_1, x_2$)-plane about the $x_3$-direction,
so that results obtained along the material axes $x_1$, $x_2$, $x_3$,
are easily transposed along the rotated axes $x^*_1$, $x^*_2$, $x_3$,
say.
On the other hand, the problem of surface Rayleigh waves is also of
prime importance because it is relevant to the study of many other
problems for anisotropic elastic half-spaces,
such as:
near-the-surface stability analysis of a deformed half-space
\cite{Biot65},
normal forces applied to a half-space \cite{Lamb04},
punch and indentation of half-space \cite{GrZe68},
steady state crack propagation \cite{Brob99}, and so on.
An up-to-date account on the research and applications of surface
acoustic waves in materials science can be found in \cite{Hess02}.
In the present paper, the \textit{secular equation} is established
for surface waves in monoclinic crystals with the plane of symmetry at
$x_3=0$, where by `secular equation' is meant the function of the
squared wave speed $c^2$ which is zero when the tractions on the plane
$x_2=0$ and at $x_2 \rightarrow \infty$ are zero.
This equation is valid for the propagation of a Rayleigh wave in
any direction of a symmetry plane for crystals possessing 
one plane of symmetry  and of course for crystals of a 
higher order symmetry class, such as orthorhombic symmetry.
It is also easy to take an eventual incompressibility of the elastic 
half-space into account \cite{DeMT02}.
Finally, after a rotation about the $x_3$-axis, the quartic secular 
equation may reduce to a biquadratic which can then be solved 
explicitly.

The secular equation is obtained in Section 3, after the equations of 
motion and boundary conditions for the problem have been recalled in
Section 2.
This equation is obtained in two different manners,
first in ``covariant'' \cite{Furs97} form, then as a quartic in the 
squared wave speed (Currie 1979, Destrade 2001, Ting 2002a).
In the final Section (\S 4), the results are applied to other
situations.
First, a rotation is made for the ($x_1,  x_2$) plane about the
$x_3$-axis and, at least for three monoclinic crystals
(diallage, gypsum, tin fluoride), two directions are found
for which the secular equation may be solved explicitly.
Then the results are specialized from monoclinic to orthorhombic
symmetry.
Finally, the constraint of incompressibility is taken into account,
and a numerical problem left open by Nair and Sotiropoulos (1999) is
resolved.

Throughout the paper, the dynamical analysis is based on  the use of
the components of the tractions rather than the displacements,
and expressions are found in terms of the stiffnesses as well as
in terms of the reduced compliances.

\section{Preliminaries}

Here we recall the equations of motion for a linearly elastic
semi-infinite body, made of a monoclinic material with  the plane of
symmetry at $x_3=0$,
and seek a solution in the form of a surface wave solution,
that is a solution which propagates in the $x_1$-direction,
leaves the plane $x_2=0$ free of tractions, and vanishes
as $x_2 \rightarrow \infty$.
Because for such materials, in-plane motions are decoupled from
anti-plane motions \cite{Stro62}, it is sufficient to seek a solution
in the form of a two-component displacement vector $\mathbf{u}$,
such as
\begin{equation} \label{u}
\mathbf{u}(x_1,x_2,t) 
 = [U_1(x_2), U_2(x_2), 0]^{\text{T}} e^{ik(x_1 -ct)},
\end{equation}
where $U_1$, and $U_2$ are functions of $x_2$ satisfying
$U_1(\infty) = U_2(\infty) = 0$, $k$ is the wave number,
and $c$ is the wave speed.

With this convention, the equations of motion are written as
\cite{Mozh95}
\begin{equation} \label{syst-u}
\mbox{\boldmath $\alpha$} \mathbf{U}''
 + i \mbox{\boldmath $\beta$} \mathbf{U}'
  - \mbox{\boldmath $\gamma$} \mathbf{U} = \mathbf{0},
\end{equation}
where $\mathbf{U}= [U_1,U_2]^{\text{T}}$ and the prime denotes
differentiation with respect to $kx_2$.
Here the symmetric $2 \times 2$ matrices
$\alpha_{ij}$, $\beta_{ij}$, and $\gamma_{ij}$,
are given in terms of the elastic stiffnesses $C$'s
and of the mass density $\rho$ by
\begin{equation}
\begin{array}{c}
\mbox{\boldmath $\alpha$}
= \begin{bmatrix}
 C_{66} & C_{26} \\
 C_{26} & C_{22}
\end{bmatrix},
\quad
\mbox{\boldmath $\beta$}
= \begin{bmatrix}
 2C_{16} & C_{12}+ C_{66} \\
  C_{12}+ C_{66} & 2C_{26}
\end{bmatrix},
\\
\mbox{\boldmath $\gamma$}
= \begin{bmatrix}
 C_{11} - \rho c^2 & C_{16} \\
 C_{16} & C_{66} - \rho c^2
\end{bmatrix}.
\end{array}
\end{equation}
Finally for the problem at hand, the following boundary conditions
must also be satisfied:
\begin{align}
& C_{66} U_1'(0) + C_{26}U_2'(0) + i C_{16}U_1(0) + iC_{66}U_2(0) = 0,
\nonumber \\
& C_{26} U_1'(0) + C_{22}U_2'(0) + i C_{12}U_1(0) + iC_{26}U_2(0) = 0.
\end{align}

Dual to this approach is one involving the components of the tractions
acting upon the planes parallel to the free surface, instead of the
components of the mechanical displacement.
Indeed, just as in-plane strain is decoupled from anti-plane strain,
so is in-plane stress from anti-plane stress
(Stroh 1962, Ting 1996, Destrade 2001a).
Thus, introducing the scalars functions $t_1(x_2)$ and $t_2(x_2)$,
defined by
\begin{equation} \label{sigma}
\sigma_{21}(x_1,x_2,t) = t_1(x_2)e^{ik(x_1 -ct)}, \quad
\sigma_{22}(x_1,x_2,t) = t_2(x_2)e^{ik(x_1 -ct)},
\end{equation}
where $\sigma_{21}$ and $\sigma_{22}$ are the in-plane stress
components, the equations of motions may be written as \cite{Dest01a}
\begin{equation} \label{syst-t}
\mbox{\boldmath $\widehat{\alpha}$} \mathbf{t}''
 - i \mbox{\boldmath $\widehat{\beta}$} \mathbf{t}'
  - \mbox{\boldmath $\widehat{\gamma}$} \mathbf{t} = \mathbf{0},
\end{equation}
where $\mathbf{t}= [ t_1, t_2  ]^{\mathrm{T}}$.
Here the symmetric $2\times 2$ matrices $\widehat{\alpha}_{ij}$,
$\widehat{\beta}_{ij}$, and $\widehat{\gamma}_{ij}$ are given in
terms of the components of the stiffness matrix $\mathbf{C}$
 \cite{Dest01a} or of the components of the reduced
compliance matrix $\mathbf{s'}$  \cite{Ting03} by
\begin{equation} \label{matrices}
\begin{array}{c}
\mbox{\boldmath $\widehat{\alpha}$}
= \begin{bmatrix}
 \frac{1}{\eta - X} & 0 \\
 0 & -\frac{1}{X}
\end{bmatrix},
\quad
\mbox{\boldmath $\widehat{\beta}$}
= \begin{bmatrix}
 -2\frac{r_6}{\eta -X} & \frac{1}{X} - \frac{r_2}{\eta - X} \\
   \frac{1}{X} - \frac{r_2}{\eta - X} & 0
\end{bmatrix},
\\
\mbox{\boldmath $\widehat{\gamma}$}
= \begin{bmatrix}
 n_{66} + \frac{r_6^2}{\eta -X} -\frac{1}{X}
    & n_{26} + \frac{r_2 r_6}{\eta -X} \\
 n_{26} + \frac{r_2 r_6}{\eta -X}
    & n_{22} + \frac{r_2^2}{\eta -X}
\end{bmatrix}.
\end{array}
\end{equation}
where $X = \rho c^2$ and
\begin{align} \label{coefficients}
\Delta &= \begin{vmatrix}
     C_{22} & C_{26} \\
     C_{26} & C_{66}
    \end{vmatrix}
       = C_{22} C_{66} - C_{26}^2,
\quad
\eta =  \frac{1}{\Delta}
    \begin{vmatrix}
     C_{11} & C_{12} & C_{16} \\
     C_{12} & C_{22} & C_{26} \\
     C_{16} & C_{26} & C_{66}
    \end{vmatrix}
    =   \frac{1}{s'_{11}},
\nonumber \\
r_6 &=  -\frac{1}{\Delta}\begin{vmatrix}
     C_{12} & C_{16} \\
     C_{22} & C_{26}
    \end{vmatrix} = -\frac{s'_{16}}{s'_{11}},
\quad
r_2 =  \frac{1}{\Delta}\begin{vmatrix}
     C_{12} & C_{26} \\
     C_{16} & C_{66}
    \end{vmatrix} = -\frac{s'_{12}}{s'_{11}},
\\
n_{66} &=  \frac{C_{22}}{\Delta} = \frac{1}{s'_{11}}\begin{vmatrix}
     s'_{11} & s'_{16} \\
     s'_{16} & s'_{66}
    \end{vmatrix},
\quad
n_{22} =  \frac{C_{66}}{\Delta} = \frac{1}{s'_{11}}\begin{vmatrix}
     s'_{11} & s'_{12} \\
     s'_{12} & s'_{22}
    \end{vmatrix},
\nonumber  \\
    n_{26} &=  -\frac{C_{26}}{\Delta}
 = \frac{1}{s'_{11}}\begin{vmatrix}
     s'_{11} & s'_{16} \\
     s'_{12} & s'_{26}
    \end{vmatrix}.  \nonumber
\end{align}
We recall that for two-dimensional deformations of a monoclinic
material with the plane of symmetry at $x_3=0$
involving the coordinates $x_1$ and $x_2$ only, the relevant non-zero
stiffnesses and reduced compliances are related through
\begin{equation} \label{C-s'}
 \begin{bmatrix}
  C_{11} & C_{12} & C_{16} \\
  C_{12} & C_{22} & C_{26} \\
  C_{16} & C_{26} & C_{66} \end{bmatrix}
 \begin{bmatrix}
  s'_{11} & s'_{12} &  s'_{16} \\
  s'_{12} & s'_{22} &  s'_{26} \\
  s'_{16} & s'_{26} &  s'_{66}
  \end{bmatrix}
  = \begin{bmatrix}
  1 & 0 & 0 \\
  0 & 1 & 0 \\
  0 & 0 & 1
  \end{bmatrix}.
\end{equation}

Finally the boundary conditions are written in a much simpler form than
when displacement components are involved, as
\begin{equation} \label{BC}
t_1(0) = t_2(0) = 0, \quad \text{and} \quad
 t_1(\infty) = t_2(\infty) =0.
\end{equation}

\section{The secular equation}

\subsection{The characteristic polynomial}

Now we seek solutions to the equations of motion \eqref{syst-t}
in the form
\begin{equation}
\mathbf{t}(x_2) = e^{ikp x_2} \mathbf{T},
\end{equation}
where $\Im(p) > 0$, to ensure the decay of the wave amplitude away
from the free surface, and $\mathbf{T}$ is a constant vector.
So we have by \eqref{syst-t},
\begin{equation} \label{motion}
 \begin{bmatrix}
-\widehat{\alpha}_{11} p^2  + \widehat{\beta}_{11} p
- \widehat{\gamma}_{11}
     & \widehat{\beta}_{12} p - \widehat{\gamma}_{12} \\
 \widehat{\beta}_{12} p - \widehat{\gamma}_{12}
         & -\widehat{\alpha}_{22} p^2 - \widehat{\gamma}_{22}
 \end{bmatrix}
\mathbf{T}=
 \begin{bmatrix}
 0 \\ 0
 \end{bmatrix}.
\end{equation}
Hence, for nontrivial solutions to exist, $p$ must be the root of a
quartic, which corresponds to the determinant of the matrix above
being equal to zero.
This quartic, the \textit{characteristic polynomial} of the equations
of motion, may be written as
\begin{equation} \label{quartic-xi}
p^4 + 2\omega_3 p^3 + \omega_2 p^2 + 2 \omega_1 p
 + \omega_0 = 0,
\end{equation}
where the coefficients $\omega_3$, $\omega_2$, $\omega_1$, and
$\omega_0$ are given by
\begin{align}
& \omega_3 = -\frac{s'_{16}}{s'_{11}}, \nonumber \\
& \omega_2 = \frac{1}{s'_{11}}[ s'_{66} + 2s'_{12}
                - X(s'(1,2) + s'(1,6))], \nonumber  \\
& \omega_1 = -\frac{1}{s'_{11}}[ s'_{26} +
                  X(s'(1,2|2,6) - s'(1,6|1,2))], \\
& \omega_0 = \frac{1}{s'_{11}}[ s'_{22}
                - X(s'(1,2) + s'(2,6))
                    + X^2s'(1,2,6)]. \nonumber
\end{align}
Here, the expression $s'(n_1 \ldots n_k|m_1 \ldots m_k)$
represents the determinant of the $k \times k$ matrix which is a
submatrix of the matrix $s'_{ij}$ ($i,j=1, \ldots, 6$) and whose
components correspond to the intersections of the rows
$n_1, \ldots, n_k$ and the columns $m_1, \ldots, m_k$.
Moreover, when $n_1 = m_1, \ldots, n_k = m_k$, the shorter expression
$s'(n_1 \ldots n_k) \equiv s'(n_1 \ldots n_k| n_1 \ldots n_k)$ is
used.
The quartic \eqref{quartic-xi} was obtained by Ting (2002a, 2002b),
and by Furs (1997) in terms of invariants of the stiffness matrix
$\mathbf{C}$. Note that when $X= \rho c^2 = 0$, the quartic of the 
elastostatic case is recovered as (Steeds, 1973, p. 72)
\begin{equation}
s'_{11}p^4 - 2s'_{16}p^3 + (2s'_{12}
	+ s'_{66})p^2 - 2s'_{26}p + s'_{22} = 0.
\end{equation}

Now we use the boundary conditions \eqref{BC}$_{1,2}$ at the free
surface to establish the secular equation.
Let $p_1$ and $p_2$ be the roots of the characteristic polynomial
\eqref{quartic-xi} with positive imaginary part, and let
$\mathbf{T}^{(r)}$ be a vector satisfying \eqref{motion} when
$p = p_r (r =1,2)$.
These vectors are in the form, say,
\begin{equation} \label{T}
\mathbf{T}^{(1)} = \begin{bmatrix}
  \widehat{\alpha}_{22} p_1^2 + \widehat{\gamma}_{22}
\\ \widehat{\beta}_{12} p_1 - \widehat{\gamma}_{12}
 \end{bmatrix},
\quad
\mathbf{T}^{(2)} = \begin{bmatrix}
  \widehat{\alpha}_{22} p_2^2 + \widehat{\gamma}_{22}
\\ \widehat{\beta}_{12} p_2 - \widehat{\gamma}_{12}
 \end{bmatrix}.
\end{equation}
Then, assuming $p_1 \ne p_2$, 
the tractions $\mathbf{t}$ defined in \eqref{sigma}
are a combination of $\mathbf{T}^{(1)}$
and  $\mathbf{T}^{(2)}$ for some constants $q_1$ and $q_2$,
\begin{equation} \label{t-T}
\mathbf{t}(x_2) = q_1 e^{ikp_1 x_2} \mathbf{T}^{(1)}
            + q_2 e^{ikp_2 x_2} \mathbf{T}^{(2)}.
\end{equation}
Using Ting's (2002a, 2002b) notation, $\mathbf{t}$ may
be written as
\begin{equation} 
\mathbf{t}(x_2) = \mathbf{B} <e^{i k x_*}> \mathbf{q},
\end{equation}
where $\mathbf{q}$ is the vector $[q_1, q_2]^{\text{T}}$, and
the matrices $\mathbf{B}$ and $ <e^{i k x_*}> $ are defined by
\begin{equation}
\mathbf{B} = [\mathbf{T}^{(1)},  \mathbf{T}^{(2)}], \quad
<e^{i k x_*}> = \text{diag} (e^{i k p_1 x_2}, e^{i k p_2 x_2}) .
\end{equation}
The tractions satisfy the boundary conditions \eqref{BC}$_1$, that is
$\mathbf{t}(0)=\mathbf{0}$, when
\begin{equation}
\mathbf{B} \mathbf{q} = \mathbf{0}.
\end{equation}
This system has non trivial solutions when $\text{det} \mathbf{B} =0$, 
that is when the following  \textit{secular equation} is satisfied,
\begin{equation} \label{seculComplex}
\widehat{\alpha}_{22}\widehat{\beta}_{12} p_1p_2
    - \widehat{\alpha}_{22}\widehat{\gamma}_{12} (p_1 + p_2)
- \widehat{\beta}_{12}\widehat{\gamma}_{22}=0.
\end{equation}

Now we try to obtain more satisfactory expressions for this
equation.

\subsection{The ``covariant'' secular equation}

First, we decompose $p_1 +p_2$ and  $p_1p_2$ into their
real and imaginary parts as
\begin{equation}
 p_1 + p_2 = u^+ + iu^-, \quad
 p_1p_2 = v^+ + iv^-.
\end{equation}
It is known that when the roots of the quartic \eqref{quartic-xi}
are $p_1$, $p_2$, $\overline{p}_1$, and $\overline{p}_2$,
then $u^+$, $u^-$, $v^+$, and $v^-$ satisfy
\begin{align} \label{omega-u-v}
&\omega_3 = - u^+,
 & \omega_2 = (u^+)^2 + (u^-)^2 + 2v^+, \nonumber \\
&\omega_1 = - u^+ v^+ - u^-v^-,
 & \omega_0 = (v^+)^2 + (v^-)^2,
\end{align}
which leads to the following cubic for $v^+$,
\begin{equation} \label{cubic}
(v^+)^3 + b_2 (v^+)^2 + b_1 (v^+) + b_0 = 0,
\end{equation}
where $b_2 = -\omega_2/2$, $b_1 = \omega_1 \omega_3 - \omega_0$,
and $b_0 = [\omega_0(\omega_2 -\omega_3^2) - \omega_1^2]/2$.
On the other hand, the secular equation \eqref{seculComplex}
may also be separated into its real and imaginary parts,
\begin{equation} \label{seculReIm}
\widehat{\alpha}_{22}\widehat{\beta}_{12} (v^+)
    - \widehat{\alpha}_{22}\widehat{\gamma}_{12} (u^+)
- \widehat{\beta}_{12}\widehat{\gamma}_{12}=0,
\quad
\widehat{\alpha}_{22}\widehat{\beta}_{12} (v^-)
    - \widehat{\alpha}_{22}\widehat{\gamma}_{12} (u^-) =0.
\end{equation}

At this point, it is important to emphasize that the system of
six equations \eqref{omega-u-v} and \eqref{seculReIm} for the five
unknowns $u^+$, $u^-$, $v^+$, $v^-$, and $X$, is consistent.
This is due a fundamental result of the modern
theory of surface waves in anisotropic elasticity by Stroh (1962),
which states that the complex secular equation
\eqref{seculComplex} is actually equivalent to a single real
equation (for alternative proofs see Currie, 1974; or Taylor, 1978).
In fact, it can be proved that with the appropriate normalization
of the Stroh eigenvectors, the real and imaginary parts of the
secular equation are equal but opposite in sign \cite[p.469]{Ting96}.

Now,  $v^+$ must satisfy the cubic \eqref{cubic} and
also the linear equation \eqref{seculReIm}$_1$,
which using \eqref{omega-u-v}$_1$, is
\begin{multline} \label{v+}
(v^+) + a_0 = 0, \\ \text{where} \quad
a_0 = X \frac{s'_{16}[s'_{26}- Xs'(1,6|1,2)]}
        {s'_{11}[1 - X(s'_{11} - s'_{12})]}
      + X \frac{s'_{22} - s'(1,2)}{1 - Xs'_{11}}.
\end{multline}
Working along similar lines but with the displacements components
\eqref{u} rather than with the traction components \eqref{sigma},
Furs (1997) showed that $v^+$ was simultaneously the root of
a cubic (as in \eqref{cubic}) and of a quadratic (in contrast to
\eqref{v+}).
By writing the resultant of those two polynomials, he obtained
the secular equation in ``covariant form'' (his wording) as
corresponding to the nullity of a $5 \times 5$ determinant.
The same approach applied here to \eqref{cubic} and \eqref{v+},
would yield the ``covariant'' secular equation as corresponding to the
nullity of a $4 \times 4$ determinant.
However, $v^+$ is easily deduced from \eqref{v+} as $v^+ = -a_0$
and substituted into \eqref{cubic} to yield the
\textit{``covariant'' secular equation},
\begin{equation} \label{seculCov}
a_0^3 - b_2 a_0^2 + b_1 a_0 - b_0 = 0.
\end{equation}
Although not stated as such, Furs's ``covariant'' secular equation
is a polynomial of degree 6 in $X=\rho c^2$.
Equation \eqref{seculCov} is a polynomial of degree 9 in
$X=\rho c^2$, but it may be factorized into the product of two
polynomials, one of degree 5, which corresponds to spurious roots,
and one of degree 4, which is the \textit{quartic secular equation}.
This very equation is obtained  in a different and more direct
manner in the next subsection as Eq.\eqref{quartic}.

\subsection{The quartic secular equation}

In order to obtain the quartic secular equation directly, we recall
that the vectors $\mathbf{T}^{(1)}, \mathbf{T}^{(2)}$, in \eqref{T} 
were computed using the second line of \eqref{motion}.
When the first line is used, these vectors are
\begin{equation} \label{T2nd}
\mathbf{T}^{(1)} = \begin{bmatrix}
  \widehat{\beta}_{12} p_1 - \widehat{\gamma}_{12}
\\ \widehat{\alpha}_{11} p_1^2 - \widehat{\beta}_{11} p_1 
 + \widehat{\gamma}_{11}
 \end{bmatrix},
\quad
\mathbf{T}^{(2)} = \begin{bmatrix}
 \widehat{\beta}_{12} p_2 - \widehat{\gamma}_{12}
\\ \widehat{\alpha}_{11} p_2^2 - \widehat{\beta}_{11} p_2
 + \widehat{\gamma}_{11}
 \end{bmatrix}.
\end{equation}
By the same steps that lead from \eqref{t-T} to \eqref{seculComplex},
we obtain the following alternative form of the complex
secular equation,
\begin{equation} \label{seculComplex2}
\widehat{\alpha}_{11}\widehat{\beta}_{12} p_1p_2
    - \widehat{\alpha}_{11}\widehat{\gamma}_{12} (p_1 + p_2)
+ \widehat{\beta}_{11}\widehat{\gamma}_{12}
- \widehat{\beta}_{12}\widehat{\gamma}_{11}=0.
\end{equation}
By simple comparison of \eqref{seculComplex} and
\eqref{seculComplex2}, the complex terms involving $p_1p_2$
and $p_1 + p_2$ may be eliminated, and the secular equation
becomes real,
\begin{equation} \label{seculReal}
\widehat{\alpha}_{11} \widehat{\beta}_{12} \widehat{\gamma}_{22}
+ \widehat{\alpha}_{22}\widehat{\beta}_{11} \widehat{\gamma}_{12}
 - \widehat{\alpha}_{22} \widehat{\beta}_{12} \widehat{\gamma}_{11}=0.
\end{equation}
This equation is a quartic in $X=\rho c^2$.
It was established by this author \cite{Dest01a} in terms of the
stiffnesses and by Ting (2002a) in terms of the compliances
(see also Currie (1979) for a less explicit expression, obtained 
by writing the equations of motion for the displacements instead of
the traction components).
This polynomial of degree 4 in $X = \rho c^2$ is also the one obtained 
in the previous subsection by factorization of the ``covariant'' 
equation \eqref{seculCov}.
Note that it may be obtained directly by using an adequate combination
of the vectors \eqref{T} and \eqref{T2nd} for the colomns of the matrix
$\mathbf{B}$.
It is written explicitly with the coefficients in terms of the reduced 
compliances as
\begin{equation} \label{quartic}
d_4 X^4 + d_3 X^3 + d_2 X^2 + d_1 X -1 = 0,
\end{equation}
where
\begin{equation} \label{d's}
\begin{split}
d_4 & = s'_{11} (s'_{66} s'_{12} s'_{11}  - s'_{66} s_{11}^{'2}
    + s'_{12} s_{16}^{'2} - 2 s'_{16} s'_{11} s'_{26}
\\
& \phantom{ - s'_{66} s_{11}^{'2} + s'_{66} s'_{12} s'_{11}}
    - s'_{12} s'_{11} s'_{22} + s_{16}^{'2} s'_{11}
    + s_{12}^{'3} - s'_{11} s_{12}^{'2} + s_{11}^{'2} s'_{22}),
\\
d_3 & = - 2s_{11}^{'2} s'_{22} + s_{11}^{'3} - s_{11}^{'2} s'_{12}
    + 3s'_{66} s_{11}^{'2} + s_{12}^{'2} s'_{11}
\\
& \phantom{- 2s_{11}^{'2} s'_{22} + s_{11}^{'3}}
    - 2s_{16}^{'2} s'_{11} + 4 s'_{16} s'_{11} s'_{26}
    - 2 s'_{66} s'_{11} s'_{12} + s'_{22} s'_{11} s'_{12}
    - s_{16}^{'2} s'_{12},
\\
d_2 & = - 3s'_{11} s'_{66} + s_{16}^{'2} - 2s'_{16} s'_{26}
    - 3s_{11}^{'2} + s'_{11} s'_{22} + 2s'_{11}s'_{12}
    + s'_{66} s'_{12},
\\
d_1 & = 3s'_{11} - s'_{12} + s'_{66}.
\end{split}
\end{equation}
Values of the Rayleigh wave speed $c_R$ are given for the 12
crystals of Table 1 in \cite{Dest01a}.
They correspond to the square root of the least positive root
of \eqref{quartic}.

In the isotropic case, the reduced compliances take the following
values,
\begin{equation}
s'_{11}=s'_{22}= \frac{c_P^2}{4 \rho c_S^2(c_p^2-c_S^2)},
\quad
s'_{12}= \frac{2c_S^2 - c_P^2}{4\rho c_S^2(c_p^2-c_S^2)},
\quad
s'_{66}= \frac{1}{\rho c_S^2},
\end{equation}
and $ s'_{16}=s'_{26}=0$,
where $c_P$ and $c_S$ are the speeds of the longitudinal and
transverse bulk waves, respectively.
Then the quartic \eqref{quartic} factorizes into the product of a
polynomial of degree one and of the cubic found by Rayleigh (1885),
\begin{equation} \label{RayleighCubic}
\big{(}\frac{c^2}{c_S^2}\Big{)}^3
 - 8\Big{(}\frac{c^2}{c_S^2}\Big{)}^2
   + \Big{(}24 - \frac{16c_S^2}{c_P^2}\Big{)}\frac{c^2}{c_S^2}
     - 16\Big{(}1 - \frac{c_S^2}{c_P^2}\Big{)} =0.
\end{equation}

\section{Applications}

Now we apply the results of the previous section to other settings,
namely, to the case of a surface wave propagating in any direction
in the plane of symmetry $x_3=0$ for monoclinic crystals;
then  to the case of a surface wave propagating along a material axis
for rhombic crystals;
and finally to the case where the half-space is made of an
incompressible monoclinic material.
We also seek exact analytic solutions for the speed of Rayleigh waves.

\subsection{Rotation in the plane of symmetry and explicit wave speeds}

Monoclinic materials with the plane of symmetry at $x_3=0$ have
a stiffness matrix $\mathbf{C}$ which is `structurally invariant'
that is, a matrix whose components which are zero remain zero after
a rotation of the coordinate axes around the $x_3$-axis \cite{Bond43}.
Ting (2000) proved recently that the submatrix of the
reduced compliance matrix appearing in \eqref{C-s'} is also
structurally invariant.
Here, we exploit this property in order to derive the secular equation
for surface waves propagating in any direction in the plane of
symmetry.
Previous efforts covering this topic include those of Chadwick and
Wilson (1992) and of Furs (1997).

First, we consider a surface wave propagating on the plane
$x_2^*=0$ and polarized in the $x_1^*$-direction,
where the coordinate system $x_i^*$ is obtained from the material
axes coordinate system $x_i$ through a rotation about the $x_3$-axis
by an arbitrary angle $\theta$, say.
Hence,
\begin{equation}
\begin{bmatrix}
x_1^*\\
x_2^*\\
x_3^*
\end{bmatrix} =
\begin{bmatrix}
 m  & n & 0 \\
-n  & m & 0 \\
 0  & 0 & 1
\end{bmatrix}   \begin{bmatrix}
            x_1    \\
            x_2    \\
            x_3
        \end{bmatrix}
,\quad m = \cos \theta, \quad n = \sin \theta.
\end{equation}

Next, following Ting (2000), we infer that all the results
from the previous section are directly applicable to this wave,
as long as the reduced compliances $s'_{ij}$, ($i,j=1,2,6$) are
replaced by the following `starred' quantities,
\begin{align} \label{s'rotated}
& (s'_{11})^* = s'_{11}m^4 + (2s'_{12} + s'_{66})m^2 n^2 + s'_{22}n^4
        + 2(s'_{16}m^2 + s'_{26}n^2)mn,
\nonumber \\
& (s'_{22})^* = s'_{22}m^4 + (2s'_{12} + s'_{66})m^2n^2 + s'_{11}n^4
        - 2(s'_{26}m^2 + s'_{16}n^2)mn,
\nonumber \\
& (s'_{12})^* = s'_{12}
    + (s'_{11} + s'_{22} - 2s'_{12} - s'_{66})m^2n^2
    -  (s'_{16} - s'_{26})(m^2 - n^2)mn,
\\
& (s'_{16})^* = s'_{16}m^4 - s'_{26}n^4 -3(s'_{16}-s'_{26})m^2 n^2
\nonumber \\
& \phantom{(s'_{16})^* = s'_{16}m^4 -}
        - [2s'_{11}m^2  - 2s'_{22}n^2
            - (2s'_{12} + s'_{66})(m^2 - n^2)]mn,
\nonumber \\
& (s'_{26})^* = s'_{26}m^4 - s'_{16}n^4 + 3(s'_{16}-s'_{26})m^2 n^2
\nonumber \\
& \phantom{(s'_{26})^* = s'_{26}m^4 -}
        + [2s'_{22}m^2  - 2s'_{11}n^2
            - (2s'_{12} + s'_{66})(m^2 - n^2)]mn,
\nonumber \\
& (s'_{66})^* = s'_{66}
    + 4(s'_{11} + s'_{22} - 2s'_{12} - s'_{66})m^2n^2
    -  4(s'_{16} - s'_{26})(m^2 - n^2)mn.
\nonumber
\end{align}
In particular, the secular equation for the surface wave is the
starred version of \eqref{quartic}, that is
$d_4^* X^4 + d_3^* X^3 + d_2^* X^2 + d_1^* X -1 = 0$.

Because the coefficients $d^*_4$, $d^*_3$, $d^*_2$, and $d^*_1$ in
this quartic are functions of $\theta$, it might be possible that for
certain angles, the quartic turns into a biquadratic, for which
the real root $X$ may be found explicitly.
Some work has been devoted to the search of explicit expressions
for the speed of elastic surface waves.
For instance, Lamb (1904) noted that Rayleigh's cubic
equation \eqref{cubic} factorizes into the product of a polynomial
of degree one and of a quadratic in $c^2/c_S^2$ when Poisson's ratio
is $\textstyle{\frac{1}{4}}$, that is when $c_P^2 = 3c_S^2$ or
equivalently, when the two Lam\'e constants are equal;
in that case, $c_R^2 = 2c_S^2(1-1/\sqrt 3)$.
In the general case, closed-form expressions are rather cumbersome
for the relevant root of the cubic in isotropic \cite{Nkem97} or
orthorhombic \cite{Rome01} half-spaces and only approximate expressions
are sought  (Royer and Dieulesaint 1984, Mozhaev 1991).
However, Mozhaev (1995) showed that for the special
orthorhombic materials such that $c_{12}=c_{66}$ (or with an
equivalent relationship for different choices of the material
axes), the squared Rayleigh wave speed could be obtained as the root
of a quadratic.
Similarly, Ting (2002b) showed that the quartic
\eqref{quartic} simplifies to the product of a squared polynomial
of degree one and of quadratic in $X$ for the special monoclinic
materials with the plane of symmetry at $x_3=0$ such that
$s'_{16}-2s'_{26}=s'_{12}=0$.
Now the starred version of the quartic \eqref{quartic} may be
rewritten in canonical form,
\begin{equation}
 Y^4 + a Y^2 + b Y + e = 0,
\quad
Y = X + \frac{d^*_3}{4d^*_4},
\end{equation}
where
\begin{align}
a & = [8 d^*_2d^*_4 - 3 (d^*_3)^2]/[8(d^*_4)^2],
\nonumber \\
b & =
  [(d^*_3)^3 - 4 d^*_2 d^*_3 d^*_4 + 8 d^*_1(d^*_4)^2]/[8(d^*_4)^3],
\\
e & = [16 d^*_2(d^*_3)^2 d^*_4 - 3(d^*_3)^4 - 256(d^*_4)^3
    - 64 d^*_1 d^*_3 (d^*_4)^2]/[256(d^*_4)^4].
\nonumber
\end{align}
Clearly, it becomes a biquadratic if $b=0$ for a certain angle
$\theta = \alpha$ say.
Then, solving the equation $b=0$ at $\theta = \alpha$ for
$d_2^\alpha$, for instance, yields the following biquadratic,
\begin{equation} \label{biquadratic}
 Y^4 + 2\Big{[}\frac{d^\alpha_1}{d^\alpha_3}
     - \big{(}\frac{d^\alpha_3}{4d^\alpha_4}\big{)}^2\Big{]}Y^2
 + \Big{(}\frac{d^\alpha_3}{4d^\alpha_4}\Big{)}^4
  -  \frac{d^\alpha_1d^\alpha_3}{8(d^\alpha_4)^2}
    - \frac{1}{d^\alpha_4}= 0,
\end{equation}
whose explicit relevant root
$X = \rho c_R^2 = Y - d_3^\alpha / (4 d_4^\alpha)$ is
\begin{equation} \label{explicitX}
X = -  \frac{d^\alpha_3}{4d^\alpha_4} +
 \sqrt{\big{(}\frac{d^\alpha_3}{4d^\alpha_4}\big{)}^2
             - \frac{d^\alpha_1}{d^\alpha_3}
   - \sqrt{\big{(}\frac{d^\alpha_1}{d^\alpha_3}\big{)}^2
            + \frac{1}{d^\alpha_4}}}.
\end{equation}
Twice, the resolution allowed for a plus or a minus sign.
One time, the plus sign was selected because $X=\rho c_R^2$ must be
positive;
the other time, the minus sign was selected by continuity with the
known result \cite{Lamb04} in the special isotropic case where
Poisson's ratio is $\textstyle{\frac{1}{4}}$.

Of course, the existence of an angle $\theta = \alpha$ such that
$b=0$ is not guaranteed.
However, numerical simulations show that at least 
for diallage, gypsum, and tin fluoride, two angles $\alpha$ may 
indeed be found such that a
surface wave propagating in the $x_1^\alpha$-direction with
attenuation in the $x_2^\alpha$-direction, where
($x_1^\alpha, x_2^\alpha$) are obtained from the material axes
($x_1,x_2$) by a rotation about the $x_3$-axis of the angle $\alpha$,
has a velocity $c_R = \sqrt{X/\rho}$ which is given explicitly by
\eqref{explicitX}.
For diallage, the angles are $\alpha = 80.24^o$ and $87.64^o$ with
corresponding Rayleigh speeds $c_R = 3960$ and $3952$ m/s,
respectively;
for gypsum, the angles are $\alpha = 18.82^o$ and $65.00^o$ with
corresponding Rayleigh speeds $c_R = 2895$ and $2946$ m/s,
respectively;
and for tin fluoride, the angles are $\alpha = 4.01^o$ and $31.21^o$
with corresponding Rayleigh speeds $c_R = 1324$ and $1351$ m/s,
respectively.

\subsection{Orthorhombic crystals}

For orthorhombic crystals, $s'_{16}=s'_{26}=0$,
so that $\widehat{\beta}_{11} = \widehat{\gamma}_{12} =0$
and the results obtained for monoclinic materials are greatly
simplified.
In particular, the characteristic polynomial \eqref{quartic-xi} is
now a biquadratic in $p$,
\begin{equation} \label{biquadratic-xi}
p^4 - S p^2 + P = 0,
\end{equation}
where the real scalars $S$ and $P$ are given by
\begin{align}
S & =  \frac{1}{s'_{11}}[ s'_{66} + 2s'_{12}
        - X(s'_{11}s'_{22} - s^{'2}_{12} + s'_{11}s'_{66})],
\nonumber  \\
P & = \frac{1}{s'_{11}}(1-Xs'_{66})[s'_{22}-X(s'_{11}s'_{22}
                        -s^{'2}_{12})].
\end{align}
Note that, depending upon the sign of $S^2-4P$ and of $S$,
the roots $p_1$ and $p_2$ of the biquadratic with positive
imaginary parts are either purely imaginary:
$p_1 = i \sqrt{(-S+ \sqrt{S^2-4P})/2}$,
$p_2 = i \sqrt{(-S- \sqrt{S^2-4P})/2}$
(when $S^2-4P>0$, $S<0$),
or of the form:
$p_1 = a + ib$, $p_2 = -a +ib$, where
$a =  \sqrt{(S + 2\sqrt{P})/4}$ and
$b =  \sqrt{(-S+ 2\sqrt{P})/4}$ (when $S^2-4P<0$).
In any case, we have
\begin{equation}    \label{xi1xi2}
p_1 p_2 = - \sqrt{P}.
\end{equation}

It is now easy to see that the \textit{secular equation} 
\eqref{seculComplex} is simplified to (using \eqref{xi1xi2})
\begin{equation} \label{secularRhombic1}
\widehat{\alpha}_{22} \sqrt{P} + \widehat{\gamma}_{22}=0.
\end{equation}

Explicitly,  the secular equation  \eqref{secularRhombic1} is
written  as
\begin{equation} \label{secularRhombic2}
(1-Xs'_{11}) \sqrt{1-Xs'_{66}}
  - X \sqrt{s'_{11}[s'_{22} -X(s'_{11}s'_{22}-s^{'2}_{12})]}
            = 0.
\end{equation}
The secular equation for surface waves in
orthorhombic crystals was first established by Skelvo (1948)
in terms of the elastic stiffnesses.
Ting (2002b) found the cubic secular equation in
terms of the elastic reduced compliances,
an equation which may be deduced from \eqref{secularRhombic2}
by rationalization;
however, the squaring process introduces spurious roots,
while the exact secular equation \eqref{secularRhombic2}
has a unique root \cite{Rome01}.

Note that  after rotation about the $x_3$-axis, the secular equation 
is again $d_4^* X^4 + d_3^* X^3 + d_2^* X^2 + d_1^* X - 1 =0$,
where the $d^*$'s are given by \eqref{d's} and \eqref{s'rotated}
with $s'_{16} = s'_{26} = 0$.
Also, special directions in which the secular equation is a
biquadratic might also be found for orthorhombic crystals,
following a procedure similar to the one exposed in the previous
subsection.

\subsection{Incompressible monoclinic materials}

According to Klintworth and Stronge (1990),
``many anisotropic composite materials appear relatively
incompressible because their bulk modulus is large compared with their
shear moduli.
In particular, low-density cellular materials are highly compliant in
shear because the flexural rigidity of the cell walls is small.''
Nair and  Sotiropoulos (1997, 1999) also studied
incompressible anisotropic materials; in particular they considered
surface waves in monoclinic materials with the plane of symmetry at
$x_3=0$, but did not obtain the secular equation explicitly.

Recently, Destrade, Martin, and Ting (2002) proved that in
linear ani\-so\-tro\-pic elasticity, the constraint of
incompressibility implied that certain relationships must be
satisfied for some compliances.
In the present context, the following relationships must hold,
\begin{equation}
s'_{11} + s'_{12} = s'_{12} + s'_{22} = s'_{16} + s'_{26} = 0,
\end{equation}
and they greatly simplify the \textit{quartic secular equation}
\eqref{quartic} to \begin{multline} \label{quarticIncompressible}
2s^{'2}_{11}(s^{'2}_{16} - s'_{11} s'_{66}) X^4
  -5s'_{11}(s^{'2}_{16} - s'_{11} s'_{66})X^3 \\
+  [3s^{'2}_{16} - 4s'_{11}(s'_{11} + s'_{66})] X^2
+ (4s'_{11} + s'_{66}) X -1 = 0.
\end{multline}
This equation was obtained by Destrade et al. (2002) in a less
explicit manner.

Nair and  Sotiropoulos (1999) introduced the constants
$\alpha$, $\beta$, and $\gamma$ defined by
\begin{equation}
\alpha = \frac{s'_{11}}{s'_{11}s'_{66}-s^{'2}_{16}},
\quad
\beta = \frac{s'_{66}}{4s'_{11}} - 1,
\quad
\gamma = -\frac{s'_{16}}{s'_{11}},
\end{equation}
When these equations are solved for $s'_{11}$, $s'_{16}$,
and $s'_{66}$, the secular equation \eqref{quarticIncompressible}
may be written as a quartic in
$x \equiv X/\alpha = \rho c^2 / \alpha$,
\begin{multline} \label{seculNaSo}
2x^4 - 5(4 \beta + 4 - \gamma^2)x^3 +
(16 \beta + 20 - 3\gamma^2)(4 \beta + 4 - \gamma^2)x^2 \\
    -4(\beta + 2)(4 \beta + 4 - \gamma^2)^2x
        +(4 \beta + 4 - \gamma^2)^3 =0.
\end{multline}
Now a numerical example is given, as the surface wave speed is
computed in the case \cite{NaSo99} where $\beta =0.3$
and $\delta = 0.1$.
Then, the secular equation \eqref{seculNaSo} is the quartic
\begin{equation}
2.0000 x^4 - 25.950 x^3 + 128.56 x^2 -247.81 x + 139.80 =0,
\end{equation}
for which the least real root is $x= 0.94671$,
in agreement with the likely limit of the `iterative
solution' given by Nair and Sotiropoulos (1999).


\end{document}